\documentstyle[12pt,psfig]{article}

\begin{document}
\title{Productions of $\sigma$ meson in $J/\psi$ and $\psi(2s)$ decays}
\author{Bing An Li\\
Department of Physics and Astronomy, University of Kentucky\\
Lexington, KY 40506, USA\\}
\maketitle
\begin{abstract}
Productions of scalar meson($\sigma$) in the decays of 
$J/\psi\rightarrow\omega+\sigma$,$J/\psi\rightarrow\gamma+\sigma$
and $\psi(2s)\rightarrow \omega+\sigma, J/\psi+\sigma, \gamma+\sigma$
and $ee^+$ annihilations in the regions of $J/\psi$ and $\psi$ are studied  
\end{abstract}
\newpage
It is known that $\sigma$ meson($f_0(600)$) is related to chiral symmetry which is
one of the fundamental features of nonperturbative QCD. On the other hand,
the identification of $\sigma$ meson is a long standing puzzle[1]. 
Because of the large decay width($\Gamma=(600-1000)MeV$[1]) it is 
very difficult to establish the $\sigma$ pole by a naive Breit-Wigner resonance[1].

In 1989 DM2 Collaboration[2] has observed a wide bump in the decays $J/\psi
\rightarrow
\omega\pi^+\pi^-, \omega\pi^0\pi^0$ at a low ($\pi\pi$) mass.
A fit to a single Breit-Wigner curve folded to a $\pi\pi$ phase space gives the 
following parameters
\begin{eqnarray}
\lefteqn{M=(414\pm20)MeV,\;\;\;\Gamma=(494\pm58)MeV,}\nonumber \\
&&BR(J/\psi\rightarrow
\omega(\pi^+\pi^-)_{low mass})=(0.16\pm0.03)\times10^{-2},\nonumber\\
&&BR(J/\psi\rightarrow\omega(\pi^0\pi^0)_{low mass})=(0.08\pm0.01\pm0.02)
\times10^{-2}.
\end{eqnarray}
In 2004 BES Collaboration[3] has reported the measurements of a large broad peak 
due to $\sigma$ meson
in $J/\psi\rightarrow\omega\pi^+\pi^-$ at low $\pi\pi$ mass. Six analysis have been 
done and the results of mass and width are presented in Table 1 and 2 of Ref.[3].

Both DM2 and BES experiments of $J/\psi\rightarrow\omega\pi\pi$ provide clear 
evidence for the existence of the scalar meson whose mass is low and width is wide.
In Refs.[4] by parametrizating the amplitudes of the two pion decays of the 
$\Upsilon(ns)$ and $\psi(2s)$ 
the mass and width of $\sigma(f_0(600))$ are determined. 

The interesting points of the $\sigma$ meson are 1)large decay width; 2)the mass 
and the decay width are in the same range. Unlike 
other particles, the determination 
of the mass and the width of the $\sigma$ meson is not an easy task[1] and 
it depends on model which is used to extract the parameters. The Table 2 of Ref.[3]
clearly shows the model dependence of the mass and the width of the $\sigma$ meson.
When Breit-Wigner formula is used the mass parameter is often taken to be a constant
and the width might depend on energy, see Ref.[3] for example.
In Ref.[5] the pole approach is used to study the $f_0(980)$. Both real
and imaginary parts of the pole are momentum dependent.

In this paper $\sigma$ meson productions in the decays $J/\psi\rightarrow\omega+\sigma
, \gamma\sigma$,
$ee^+\rightarrow J/\psi\rightarrow\omega
+\sigma$ and $\psi(2s)\rightarrow \omega+\sigma, J/\psi
+\sigma, \gamma+\sigma$ and
$ee^+\rightarrow \psi(2s)\rightarrow \omega+\sigma, J/\psi
+\sigma$ are studied by using effective Lagrangians. 
It is very interesting to notice that
both $J/\psi$ and $\psi(2s)$ are very narrow resonances and $\sigma$ is a very 
wide resonance. $J/\psi$, $\psi(2s)$, and $\omega$ are vector mesons. 
The Vector Meson Dominance(VMD) is used to predict the decay rate of
$J/\psi\rightarrow\gamma\sigma$ and to study $\psi(2s)\rightarrow\gamma\sigma$. 

The $\sigma$ meson has been observed in 
$J/\psi\rightarrow\omega(2\pi)_{low mass}$[2,3], where the low mass region is $2m_\pi$  
to $2m_K$.
Two decay modes, $J/\psi\rightarrow\omega f_2(1270), b_1(1235)\pi$ have been measured[2],  
which contribute to $J/\psi\rightarrow\omega+2\pi$ in the region of higher mass of 
two pions. The tails of the two resonances contribute to the higher
end of the mass distribution of $(2\pi)_{low mass}$ and the contributions are small. 
Estimation of the contribution 
of non-resonance to this process is needed. 
The larger branching ratio of $J/\psi\rightarrow\rho\pi$ has been 
measured to be $(1.27\pm0.09)\times10^{-2}[6]$. The coupling $\pi\omega\rho$
of the Wess-Zumino-Witten anomaly[7]
is expressed as 
\begin{equation}
{\cal L}_{\pi\omega\rho}=-\frac{N_C}{\pi^2 g^2 f_\pi}\epsilon^{\mu\nu\alpha\beta}
\partial_\mu\omega_\nu\partial_{\alpha}\rho^i_{\beta}\pi^i,
\end{equation} 
where g is the universal coupling constant in the effective meson theory
and is determined to be 0.39 by fitting $\rho\rightarrow ee^+$.
It is similar to Eq.(2) the effective Lagrangian of the coupling $J/\psi\rho\pi$
is constructed as
\begin{equation}
{\cal L}_{J/\psi\rho\pi}=g_\rho\epsilon^{\mu\nu\alpha\beta}
\partial_\mu J_\nu\partial_{\alpha}\rho^i_{\beta}\pi^i,
\end{equation}
where $g_\rho$ is a parameter which is determined by the decay rate of
$J/\psi\rightarrow\rho\pi$
\begin{equation}
\Gamma(J/\psi\rightarrow\rho\pi)=\frac{3g^2_\rho}{8\pi}\{{1\over4m^2_J}
(m^2_J+m^2_\pi-m^2_\rho)^2-m^2_\pi\}^{{3\over2}}
\end{equation}
to be
$g_\rho=1.79\times10^{-3}GeV^{-1}$. As a test of the effective Lagrangian(3),
the decay rate of $J/\psi\rightarrow\gamma\pi^0$ is calculated.
Using the VMD[8]
$\rho^0\rightarrow{1\over2}egA_\mu$ in Eq.(3),
we obtain
\begin{eqnarray}
\lefteqn{\Gamma(J/\psi\rightarrow\gamma\pi^0)=\frac{\alpha}{96}g^2_\rho g^2 m^3_J
(1-{m^2_\pi\over m^2_J})^3,}\\
&&\Gamma(J/\psi\rightarrow\gamma\pi^0)/\Gamma(J/\psi\rightarrow\rho^0\pi^0)=0.86
\times10^{-2}.
\end{eqnarray}
The experimental value of this ratio(6) is $0.93(1\pm0.45)\times10^{-2}$[6]. 
The ratio(6) is independent of $g_\rho$. Theory
agrees with data very well. The effective Lagrangian(3) and the VMD work.

Using the two vertices(2,3), the contribution of this channel, 
$J/\psi\rightarrow\rho\pi,
 \rho\rightarrow\omega\pi$ to $J/\psi\rightarrow
\omega\pi^+\pi^-$ is calculated
\begin{equation}
\Gamma(J/\psi\rightarrow\omega\pi^+\pi^-)_{\rho\pi,\rho\rightarrow\omega\pi}
=0.049keV,
\end{equation}
The experimental
value is $\Gamma(J/\psi\rightarrow\omega\pi^+\pi^-)=0.655(1\pm0.17)keV$[6].
The contribution of $J/\psi\rightarrow\rho\pi, \rho\rightarrow\omega\pi$ is 
about $7.5\%$.  
In the range of low mass of the two pions($2m_\pi-2m_K$) 
the contribution 
of this channel is about $7\%$ of the the decay rate measured by DM2[2].  
Therefore, in the low mass range the contributions of nonresonances are small and
$J/\psi\rightarrow\omega\sigma$ dominants the process of $J/\psi\rightarrow
\omega(\pi\pi)_{low mass}$.

The simplest effective Lagrangian of the coupling between $\sigma$ field and pseudoscalars is 
constructed under SU(3) symmetry
\begin{equation}
{\cal L}_{\sigma}=g_\sigma\{\pi^i\pi^i+K^+ K^- +K^0\bar{K}^0+{1\over3}
\eta\eta\}\sigma,
\end{equation}
where the coupling constant $g_\sigma$ is a parameter which can be determined by
fitting the data of $\sigma$ resonance. 
The Lagrangian(8) is used to fit the mass distribution of the two pions of the decay
$J/\psi\rightarrow\omega\pi^+\pi^-$. 

Using the Lagrangian(8), one-loop corrections of the propagator of the
$\sigma$ field are calculated. The imaginary part of the loop(pion-loop) is the decay width 
of $\sigma\rightarrow\pi\pi$ and 
a mass correction is obtained from the real part. The  couplings between
$\sigma$ meson and other particles, for example vector mesons and nucleons,
contribute to the loop diagrams too. At low energies( $4m^2_\pi
< k^2< 4m^2_K$, k is the momentum of $\sigma$ meson) 
the loop diagrams from these additional couplings
don't have imaginary part and the real part are almost constants which can be absorbed
into the bare mass of the $\sigma$ meson(see below). 

Taking one loop diagrams of pions, kaons, and $\eta$ into account, the propagator
of the $\sigma$ field is written as
\begin{equation}
\Delta(k^2)=\frac{1}{k^2-m^2_0+\Pi(k^2)}
=\frac{1}{k^2-m^2(k^2)+i\sqrt{k^2}\Gamma_\sigma(k^2)},
\end{equation}
where $m^2_0$ is the sum of the bare mass of the $\sigma$
meson and the constants separated from the loop diagram, in the energy region
$4m^2_\pi < k^2 < 4m^2_K$ $Im\Pi(k^2)$ is obtained only from the pion loop 
\begin{eqnarray}
\lefteqn{Im\Pi(k^2)=\sqrt{k^2}\Gamma_\sigma(k^2),}\nonumber \\
&&\Gamma_\sigma(k^2)=\frac{3g^2_\sigma}{8\pi}{1\over\sqrt{k^2}}
(1-{4m^2_\pi\over k^2})^{{1\over2}},
\end{eqnarray}
\begin{eqnarray}
\lefteqn{Re\Pi(k^2)_\pi=\frac{3g_\sigma^2}{8\pi^2}z_\pi log\frac{1+z_\pi}{1-z_\pi}
=\frac{g_\sigma^2}{8\pi^2}f_\pi(k^2),}\nonumber \\
&&Re\Pi(k^2)_K=\frac{g^2_\sigma}{2\pi^2}z_K atan\frac{1}{z_K}
=\frac{g^2_\sigma}{4\pi^2}f_K(k^2)\nonumber \\ 
&&Re\Pi(k^2)_\eta=\frac{g^2_\sigma}{4\pi^2}{1\over9}z_\eta atan\frac{1}{z_\eta}
=\frac{g^2_\sigma}{4\pi^2}f_\eta(k^2),\nonumber \\ 
&&z_\pi=(1-{4m^2_\pi\over k^2})^{{1\over2}},\;\;\;
z_K=({4m^2_K\over k^2}-1)^{{1\over2}},\;\;\;
z_\eta=({4m^2_\eta\over k^2}-1)^{{1\over2}},\\
&&m^2(k^2)=m^2_0+Re\Pi(k^2).
\end{eqnarray}
The numerical results of the three functions, $f_\pi(k^2), f_K(k^2), f_\eta(k^2)$
 are shown in Fig.1.
In the energy region the real part of the pion loop increases with
energy fast; the kaon loop decreases with energy slowly; the contribution of $\eta$
loop is small and it doesn't vary with $k^2$. The calculation shows that 
if the mass of the particle is at the value of $m_\rho$ or above the function is 
almost a constant in the energy region. Therefore, in this energy region the effects of heavier particle
loops can be absorbed into the mass parameter $m^2_0$. 

Usually, in a complete field theory the divergent part of this kind of loop diagram 
is used to renormalize the mass and the field. As mentioned above the 
divergent term of the loop diagram has been absorbed by the mass parameter $m_0$.
The renormalization constant of the $\sigma$ field can be defined too and it can be 
used to redefine the coupling constant $g_\sigma$ which is determined by fitting data. 
This procedure doesn't affect
the shape of the distribution of Eq(9) as a function $k^2$ and   
Eq.(9) is used to fit data. There are two parameters: $m_0$ and $g_\sigma$.

In the region of $J/\psi$ resonance besides the coupling(8)  
$J/\psi\omega\sigma$, $\gamma J/\psi$, and $\gamma\omega\sigma$ couplings contribute to
$J/\psi\rightarrow\omega\sigma, \sigma\rightarrow2\pi$ and 
$ee^+\rightarrow\omega\sigma, \sigma\rightarrow2\pi$.
It is well known that at low energies in the VMD $\rho, \omega, \phi$ mesons play dominant roles.
In Ref.[9] the VMD has been extended to $J/\psi$ meson to study 
$\eta_c\rightarrow\gamma\gamma$, $\psi(3872)\rightarrow
\gamma\sigma$ and $ee^+\rightarrow\psi(3872)+\sigma$ 
\begin{equation}
{\cal L}_{\gamma J}=eg_J\{-{1\over2}F_{\mu\nu}(\partial_\mu J_\nu-\partial_\nu 
J_\mu) +A_\mu j^{\mu}\},
\end{equation}
The Lagrangian of $J\omega\sigma$ coupling is constructed as
\begin{equation}
{\cal L}_{J\omega\sigma}=g_1(\partial_\mu J_\nu-\partial_\nu J_\mu)(\partial_\mu
\omega_\nu-\partial_\nu\omega_\mu)\sigma.
\end{equation}
Eq.(13) is the standard form of the VMD of $J/\psi$, the parameter $g_J$ is determined 
by fitting the decay rate of $J/\psi\rightarrow ee^+$, \(g_J=0.0917\).
The decay rate of $J/\psi\rightarrow\omega\sigma$ determines the coupling constant $g_1$,  
$j_\mu$ of Eq.(13) is obtained from Eq.(14) by replacing $J/\psi$ by a 
photon field
\begin{equation}
j_\mu\rightarrow eg_J A_\mu.
\end{equation}
Using this substitution(15),
the Lagrangian of the coupling $\gamma\omega\sigma$ is obtained from
the Eq.(14)
\begin{equation}
{\cal L}_{\gamma\omega\sigma}=eg_J g_1(\partial_\mu A_\nu-\partial_\nu A_\mu)
(\partial_\mu \omega_\nu-\partial_\nu\omega_\mu)\sigma.
\end{equation}
On the other hand, the Lagrangian of $J\gamma\sigma$ is obtained by using the 
VDM substitution 
\begin{equation}
\omega\rightarrow {1\over6}eg A,
\end{equation}
\begin{equation}
{\cal L}_{J\gamma\sigma}={1\over6}egg_1(\partial_\mu J_\nu-\partial_\nu J_\mu)
(\partial_\mu A_\nu-\partial_\nu A_\mu)\sigma.
\end{equation}
The expressions of Eqs.(16,18) show that current conservation is satisfied in both 
Lagrangians and Eqs.(16,18) are used to study physical processes in this paper. 
Current conservation is the reason why the effective 
Lagrangian ${\cal L}_{J\omega\sigma}$ is constructed as Eq.(14). 
Using Eqs.(14,8) it is obtained
\begin{eqnarray}
\lefteqn{\frac{d\Gamma}{dk^2}(J/\psi\rightarrow\omega\pi^+\pi^-)=\frac{g^2_1}{18\pi^2}
{1\over m^3_J}\{(m^2_J-m^2_\omega-k^2)^2-4m^2_\omega k^2\}^{{1\over2}}}\nonumber \\
&&\{m^2_J m^2_\omega+{1\over2}(m^2_J+m^2_\omega-k^2)^2\}
\frac{\sqrt{k^2}\Gamma_\sigma(k^2)}{(k^2-m^2(k^2))^2+k^2\Gamma_\sigma(k^2)},
\end{eqnarray}
where $k^2$ is the invariant mass of the two pions. 
The distribution, ${1\over g^2_1}\frac{d\Gamma}{d\sqrt{k^2}}(J/\psi\rightarrow\omega\pi^+\pi^-)$, is
shown in Fig.2.   
The two parameters, $m_0$ and
$g_\sigma$ are chosen as
\[m_0=0.35GeV,\;\;\;g_\sigma=1.31GeV,  \]
to fit
the width and mass of the $\sigma$ meson, which are determined to be
\[m_\sigma=0.579GeV,\;\;\;\Gamma_\sigma=0.460GeV.\]
The values of $m_\sigma$ and $\Gamma_\sigma$ are compatible with the DM2 and BES data.
\(\Gamma_\sigma=0.309GeV\) is obtained 
from the direct calculation of $\Gamma_\sigma$(10).  
It is different from
the width at half of the height of the $\sigma$ resonance. This is the effect of 
wide resonance.

Integrating over $k^2$, inputing
the decay rate of $J/\psi\rightarrow\omega\sigma$[2], and subtracting the background
$J/\psi\rightarrow\rho\pi, \rho\rightarrow\omega\pi$,
it is determined
\[g^2_1=0.61(1\pm0.22)10^{-6}GeV^{-2}.\]
  
Now the $\sigma(ee^+\rightarrow J/\psi\rightarrow \omega\sigma, 
\sigma\rightarrow\pi^+\pi^-)$ can be calculated. $J/\psi$ is a very narrow resonance.
Before the calculation of this cross section
we use $ee^+\rightarrow\mu\mu^+$ in the region of $J/\psi$ resonance 
to study the effect of narrow resonance.
In the literature(see Ref.[10]
for example) this process is described by one-photon exchange and $J/\psi$ resonance
($\gamma$,$\gamma-J/\psi-\gamma$)
and using Eq.(13) the cross section is expressed as
\begin{equation}
\frac{d\sigma}{dcos\theta}(ee^+\rightarrow\mu\mu^+)={\pi\alpha^2\over2}{1\over q^2}
|1+\frac{e^2 g^2_J q^2}{q^2-m^2_J+i\sqrt{q^2}\Gamma_J}|^2(1+cos^2\theta).
\end{equation}
Because $J/\psi$ is a very narrow resonance at the peak, \(q^2=m^2_J\), the value of 
the resonance 
term is very large \(\frac{e^2 g^2_J q^2}{i\sqrt{q^2}\Gamma_J}=-25.4i\equiv x\).
In field theory besides the diagrams mentioned above there are chain diagrams,
$\gamma, \gamma-J/\psi-\gamma, \gamma-J/\psi-\gamma-J/\psi-\gamma,...$. Taking
one photon propagator out a series $1+x+x^2+...$ is obtained. The value of x is large 
and perturbation at the peak of the resonance doesn't work.  
Therefore,
all the chain diagrams of photon and $J/\psi$ resonance should be taken into account
and \(1+x+x^2+...=\frac{1}{1-x}\) is obtained.
In the region of the $J/\psi$ resonance after the chain diagrams of $J/\psi$ resonance
are added up  
the photon propagator is modified to be
\begin{equation}
{1\over q^2}\{1-\frac{e^2 g^2_J q^2}{q^2-m^2_J+i\sqrt{q^2}\Gamma_J}\}^{-1}.
\end{equation}
Using Eq.(21), the cross section of $ee^+\rightarrow\mu\mu^+$
in the region of $J/\psi$ resonance is rewritten as
\begin{equation}
\frac{d\sigma}{dcos\theta}(ee^+\rightarrow\mu\mu^+)={\pi\alpha^2\over2}{1\over q^2}
|1-\frac{e^2 g^2_J q^2}{q^2-m^2_J+i\sqrt{q^2}\Gamma_J}|^{-2}(1+cos^2\theta).
\end{equation}
The comparison between theory(22) and the data[11] is shown in Fig.3. 
Theory 
agrees with data well.

Using the effective Lagrangians(13,14,18) and the modified photon propagator(21), the cross 
section
of $ee^+\rightarrow\omega\sigma, \sigma\rightarrow\pi^+\pi^-$ in the $J/\psi$ region
is obtained
\begin{equation}
\frac{d\sigma}{dk^2}={12\pi\over q^4}\frac{1}{|1-\frac{e^2 g^2_J q^2}{q^2-m^2_J
+i\sqrt{q^2}\Gamma_J}|^2}\frac{m^4_J+q^2\Gamma^2_J}{(q^2-m^2_J)^2+q^2\Gamma^2_J}
\Gamma(J/\psi\rightarrow ee^+)\frac{d\Gamma}{dk^2}(J/\psi\rightarrow\omega\sigma,\sigma
\rightarrow\pi^+\pi^-),
\end{equation}
where \(\Gamma(J/\psi\rightarrow ee^+)={4\over3}\pi\alpha^2g^2_J\sqrt{q^2}\) and
replacing $m_J$ by $\sqrt{q^2}$ in Eq.(19), $\frac{d\Gamma}{dk^2}$ is obtained.
Integrating over $k^2$ from $4m^2_\pi$ to $4m^2_K$, 
the cross section in the low mass region of $\pi\pi$ is obtained and shown in Fig.4.

The decay rate of $J/\psi\rightarrow
\gamma\sigma,\sigma\rightarrow\pi^+\pi^-$ is predicted by Eq.(18)
\begin{eqnarray}
\Gamma(J/\psi\rightarrow\gamma(\pi^+\pi^-)_{low mass})&=&\frac{\alpha g^2 
g^2_1 g^2_\sigma}{216(2\pi)^2
m^3_J}\int^{4m^2_K}_{4m^2_\pi}dk^2 \frac{(m^2_J-k^2)^3}{(q^2-m^2(k^2))^2+k^2\Gamma^2_\sigma
(k^2)}(1-{4m^2_\pi\over k^2})^{{1\over2}}\nonumber \\
B(J/\psi\rightarrow\gamma(\pi^+\pi^-)_{low mass})&=&0.19\times10^{-7}.
\end{eqnarray}
The branching ratio is very small.
This decay channel, $J/\psi\rightarrow\gamma\sigma$, has 
been measured and not been found by BES[3].
The theory is consistent with BES's search.

The productions of $\sigma$ meson in $\psi(2s)$
decays can be studied by the same theoretical approach.

$\psi(2s)$ is a narrow resonance too. In the region of $\psi(2s)$ the cross section of
$ee^+\rightarrow\mu\mu^+$ is expressed as
\begin{equation}
\sigma={4\pi\alpha^2\over3}{1\over q^2}
|1-\frac{e^2 g^2_\psi q^2}{q^2-m^2_\psi+i\sqrt{q^2}\Gamma_\psi}|^{-2},
\end{equation}
where $g_\psi$ is the coupling constant between photon and $\psi(2s)$ and is determined
to be \(g_\psi=0.051\) by fitting the decay rate of $\psi(2s)\rightarrow ee^+$.
The comparison with data[12] is shown in Fig.5.

$\psi(2s)\rightarrow\omega\pi^+\pi^-$ has been measured[13].
Like $J/\psi\rightarrow\omega\pi^+\pi^-$[2,3], the channels $ b_1\pi, 
\omega f_2(1270)$ dominate $\psi(2s)\rightarrow\omega\pi^+\pi^-$. There is small
$B(\psi(2s)\rightarrow\omega(\pi^+\pi^-)_{low mass})$, where the invariant mass of two 
pions is in the range $2m_\pi<m_{2\pi}<2m_K$. Based on the $12\%$ rule of 
the decays of $\psi(2s)$ and $J/\psi$ and using DM2 data[2], it is estimated 
\begin{equation}
B(\psi(2s)\rightarrow\omega(\pi^+\pi^-)_{low mass})=(1.92\pm0.36)\times10^{-4}.
\end{equation}
This value is consistent with BES data[13] within the large experimental errors.
The isospin of $\pi^+\pi^-$ is zero and
the branching ratio of $\psi(2s)\rightarrow\omega(\pi^0\pi^0)_{low mass}$ should be half 
of the value(26).

Comparing with $J/\psi\rightarrow\omega(\pi\pi)_{low mass}$, 
it is expected that
$\psi(2s)\rightarrow\omega\sigma, \sigma\rightarrow2\pi$ 
is the main source of $\psi(2s)\rightarrow\omega(\pi^+\pi^-)_{low mass}$. In Eq.(19)
replacing $m_J$ and $g_1$ by $m_\psi$ and $g_1(2s)$ respectively, 
the distribution ${1\over g^2_1(2s)}\frac{d\Gamma}{dk^2}
(\psi(2s)\rightarrow
\omega\sigma,\sigma\rightarrow\pi^+\pi^-)$ is shown in Fig.6.  
$g_1(2s)$ is determined by the branching ratio (26) to be   
\(g_1(2s)=0.37\times10^{-3}GeV^{-1}\). 
It is similar to Eq.(23), 
in $\psi(2s)$ region $\sigma(ee^+\rightarrow\omega\pi^+\pi^-)$ is obtained
\begin{equation}
\frac{d\sigma}{dk^2}={12\pi\over q^4}\frac{1}{|1-\frac{e^2 g^2_\psi q^2}{q^2-m^2_\psi
+i\sqrt{q^2}\Gamma_\psi}|^2}\frac{m^4_\psi+q^2\Gamma^2_\psi}{(q^2-m^2_\psi)^2
+q^2\Gamma^2_\psi}
\Gamma(\psi\rightarrow ee^+)\frac{d\Gamma}{dk^2}(\psi\rightarrow\omega\sigma,\sigma
\rightarrow\pi^+\pi^-),
\end{equation}
The results are shown in Fig.7. $\sigma(ee^+\rightarrow\omega\pi^+\pi^-)$
in $\psi(2s)$ region is much smaller than $\sigma(ee^+\rightarrow\omega\pi^+\pi^-)$ 
in $J/\psi$ region. 

The decays $\psi(2s)\rightarrow J/\psi\pi^+\pi^-, J/\psi\pi^0\pi^0$ have been measured
[12,14]. The ratio of the two decay rates is about 2. The isospin of two pions 
is zero. The distribution of the invariant mass of two pion shows a bump in the
kinematic region. There are many theoretical studies on the $\pi\pi$ decay channels[15].
It is interesting to notice that the bump of the invariant mass of two pions 
is just the region of the mass of $\sigma$ meson. Therefore, it is possible that 
$\sigma$ meson is produced in the decay
$\psi(2s)\rightarrow J/\psi2\pi$[4]. 
In this paper effective Lagrangian has been used to study the $\sigma$ productions in $J/\psi
,\psi(2s)\rightarrow\omega\pi\pi$. The mass and the width of $\sigma$ meson determined 
are consistent
with experimental fits. The prediction of $J/\psi\rightarrow\gamma\sigma$ agrees with BES's
search. Now the same method can be used to study $\psi(2s)\rightarrow J/\psi+\sigma, 
\sigma\rightarrow\pi\pi$.

It is similar to Eqs.(13,14) the effective Lagrangians related to $\psi(2s)$
are constructed as
\begin{eqnarray}
\lefteqn{{\cal L}_{\gamma \psi}=eg_\psi\{-{1\over2}F_{\mu\nu}(\partial_\mu \psi_\nu
-\partial_\nu
\psi_\mu) +A_\mu j^{\mu}\},}\\
&&{\cal L}_{\psi J\sigma}=g_2(\partial_\mu \psi_\nu-\partial_\nu \psi_\mu)(\partial_\mu
J_\nu-\partial_\nu J_\mu)\sigma,
\end{eqnarray}
where $j_\mu$ is obtained from Eq.(29) by the substitution \(\psi(2s)\rightarrow eg_\psi 
A_\mu\). Replacing $g_1, m_J$, and $m_\omega$ by $g_2, m_\psi$, and $m_J$ in Eq.(19) 
respectively,
the distribution ${1\over g^2_2}\frac{d\Gamma}{dk^2}(\psi\rightarrow J/\psi\pi^+\pi^-)$ 
is obtained and
is 
plotted in Fig.8. which shows that the distribution 
${1\over g^2_2}\frac{d\Gamma}
{d\sqrt{k^2}}$ is pretty flat in the range of $\sqrt{k^2}<0.5GeV$. This behavior is 
inconsistent with BES's data(see Fig.6 of Ref.[14]). In this theoretical approach
if we adjust the decay width of $\sigma$ to a lower value(about 200 MeV)
(reduce the parameter g from
0.065 to 0.03) the distribution showed in Fig.6 of.Ref.[14] can be fitted very well.
However, the narrow $\sigma$-resonance cannot fit the data pf $J/\psi\rightarrow
\omega(\pi\pi)_{low mass}$ and inconsistent with all other findings(wide $\sigma$ resonance).
Therefore, from this theoretical approach we can only say that 
the decays $\psi(2s)\rightarrow J/\psi+\sigma$ might be part of 
$\psi(2s)\rightarrow J/\psi\pi\pi$ and it might not be the main source of $\psi(2s)\rightarrow 
J/\psi+\pi\pi$. We cannot predict the branch ratio of  
$\psi(2s)\rightarrow J/\psi+\sigma$. 

The phase space of $\psi\rightarrow J/\psi\pi^+\pi^-$ is limited. Measurements
of $\psi(2s)\rightarrow\gamma(\pi\pi)_{low mass}$ can provide useful information about
the decay channel $\psi(2s)\rightarrow J/\psi+\sigma$.
The decay rate of $\psi\rightarrow\gamma\pi^+\pi^-$ has been measured[16] 
in the range of $m_{\pi\pi}>0.9GeV$. 
The distribution of the invariant mass of two pions
of $\psi\rightarrow\gamma+\sigma, \sigma\rightarrow\pi^+\pi^-$ can be predicted 
by this approach of 
effective Lagrangian.
Using the VMD of $J/\psi$[9], it is obtained from Eq.(29)
\begin{equation}
{\cal L}_{\psi\gamma\sigma}=eg_2 g_J(\partial_\mu\psi_\nu-\partial_\nu\psi_\mu)
(\partial_\mu A_\nu-\partial_\nu A_\mu)\sigma.
\end{equation}
There is another term which is obtained from ${\cal L}_{\psi\omega\sigma}$ by the substitution
$\omega\rightarrow{1\over 6}g A$. Using the values of g and g(2s), the contribution of this
new term is negligible.
From Eqs.(30,8) it is obtained
\begin{eqnarray}
\Gamma(\psi\rightarrow\gamma(\pi^+\pi^-)_{low mass})&=&\frac{\alpha g^2_J
g^2_2 g^2_\sigma}{6(2\pi)^2
m^3_J}\int^{4m^2_K}_{4m^2_\pi}dk^2 \frac{(m^2_\psi-k^2)^3}{(q^2-m^2(k^2))^2+k^2\Gamma^2_\sigma
(k^2)}(1-{4m^2_\pi\over k^2})^{{1\over2}}.
\end{eqnarray}
The distribution ${1\over g^2_2}\frac{d\Gamma}{d\sqrt{k^2}}$ of $\psi(2s)\rightarrow\gamma\sigma,
\sigma\rightarrow (2\pi)_{low mass}$ is predicted(Fig.9), 
where $2m_\pi<m(2\pi)<2m_K$. The mass and the width determined from this process are the same 
as the ones determined from $J/\psi\rightarrow\omega(\pi\pi)_{lowmass}$.
We cannot predict the decay rate. 

In summary, the effective Lagrangians and the VMD are used to study the productions of 
the $\sigma$ meson in the decays of $J/\psi$ and $\psi(2s)$ and $ee^+$ annihilations.
The decay rates of $J/\psi\rightarrow\gamma\sigma$ and $\psi(2s)\rightarrow\omega\sigma$
are predicted. $\psi(2s)\rightarrow J/\psi\sigma$ might not be the dominant process
for $\psi(2s)\rightarrow J/\psi\pi\pi$.

The author likes to than X.H.Mo for help. This study is supported by a DOE grant.

\pagebreak
\begin{flushleft}
{\bf Figure Captions}
\end{flushleft}
{\bf Fig. 1} Real parts of the pion(top), kaon, and $\eta$ loops(bottom).

{\bf Fig. 2} Distribution $\frac{d\Gamma}{d\sqrt{k^2}}
(J/\psi\rightarrow\omega\pi^+\pi^-)$.

{\bf Fig.3} $\sigma(ee^+\rightarrow\mu\mu^+)$ in the region of $J/\psi$, $cos\theta<0.6$.

{\bf Fig.4} $\sigma(ee^+\rightarrow\omega+\pi^+\pi^-)$ in the region of $J/\psi$, 
$4m^2_\pi<k^2<4m^2_K$.

{\bf Fig.5} $\sigma(ee^+\rightarrow\mu\mu^+)$ in the region of $\psi(2s)$.

{\bf Fig.6} Distribution ${1\over g^2_1(2s)}\frac{d\Gamma}{d\sqrt{k^2}}
(\psi(2s)\rightarrow\omega\pi^+\pi^-)$.

{\bf Fig.7} $\sigma(ee^+\rightarrow\omega+\pi^+\pi^-)$ in the region of $\psi(2s)$,
$4m^2_\pi<k^2<4m^2_K$.

{\bf Fig.8} Distribution ${1\over g^2_2}\frac{d\Gamma}{d\sqrt{k^2}}
(\psi(2s)\rightarrow J/\psi+\pi^+\pi^-)$.

{\bf Fig.9} Distribution ${1\over g^2_2}\frac{d\Gamma}{d\sqrt{k^2}}(\psi(2s)\rightarrow 
\gamma\sigma,\sigma\rightarrow\pi^+\pi^-)$.

\begin{figure}
\begin{center}
\psfig{figure=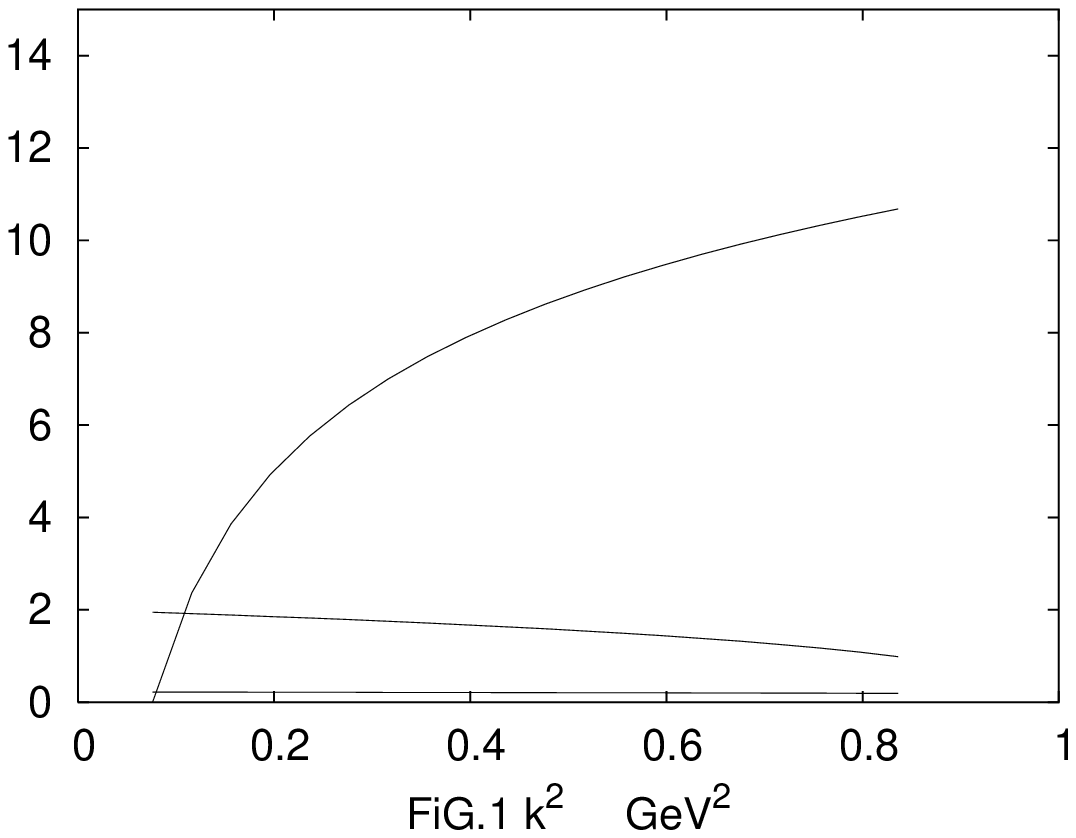}
FIG. 1.
\end{center}
\end{figure}
\begin{figure}
\begin{center}
\psfig{figure=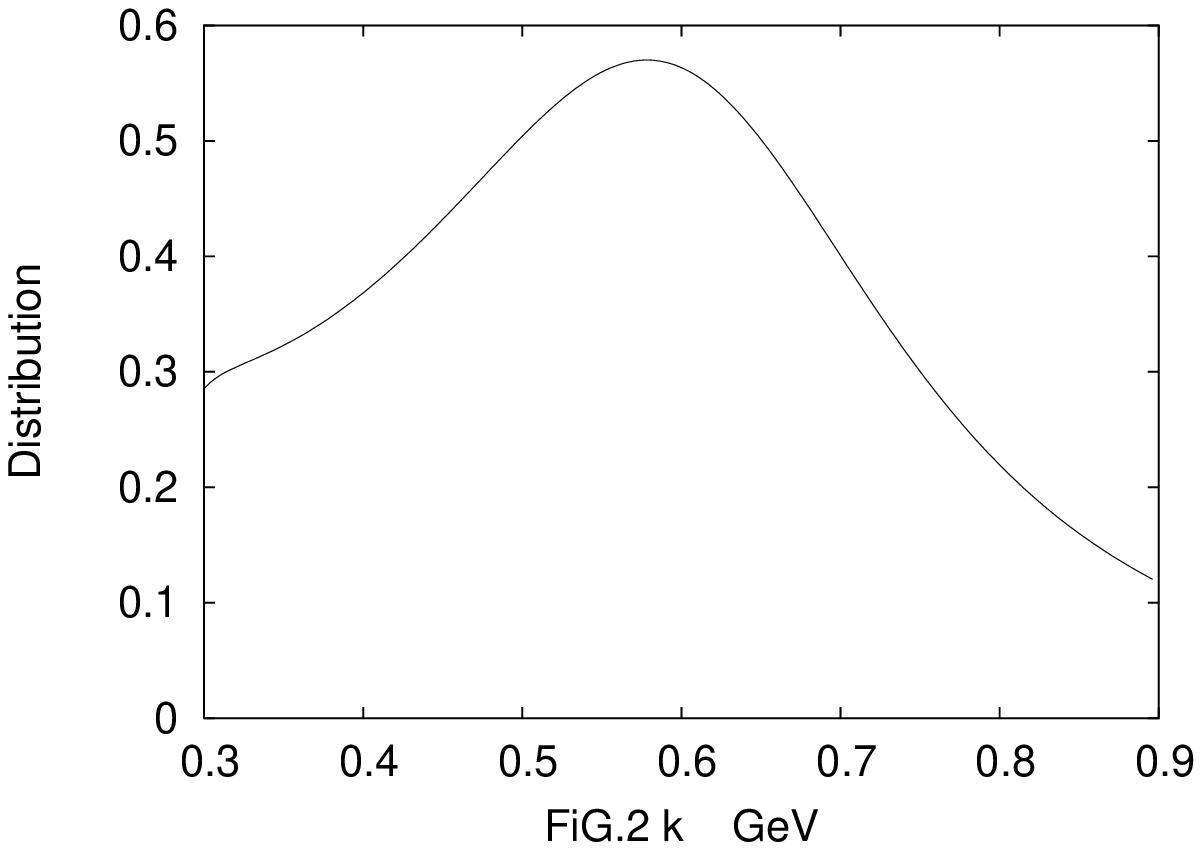}
FIG. 2.
\end{center}
\end{figure}
\begin{figure}
\begin{center}
\psfig{figure=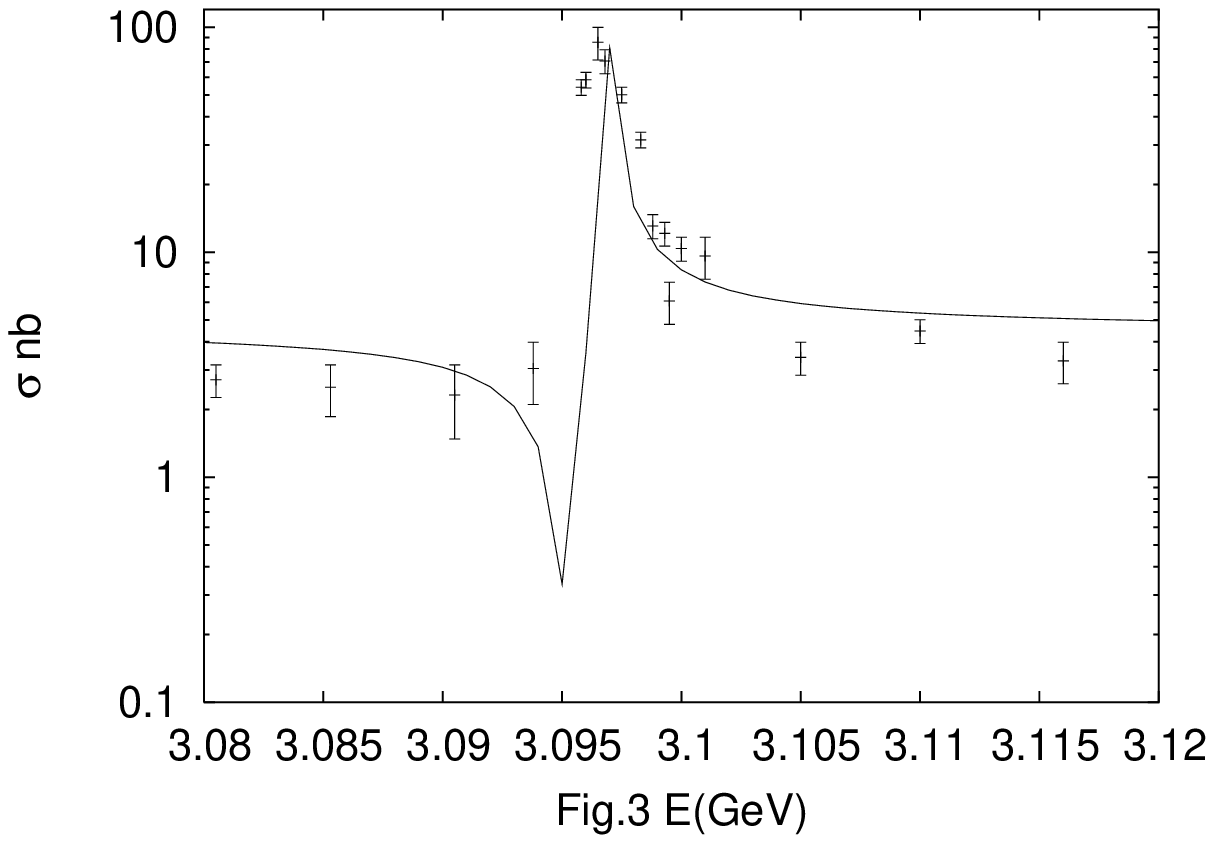}
FIG. 3.
\end{center}
\end{figure}
\begin{figure}
\begin{center}
\psfig{figure=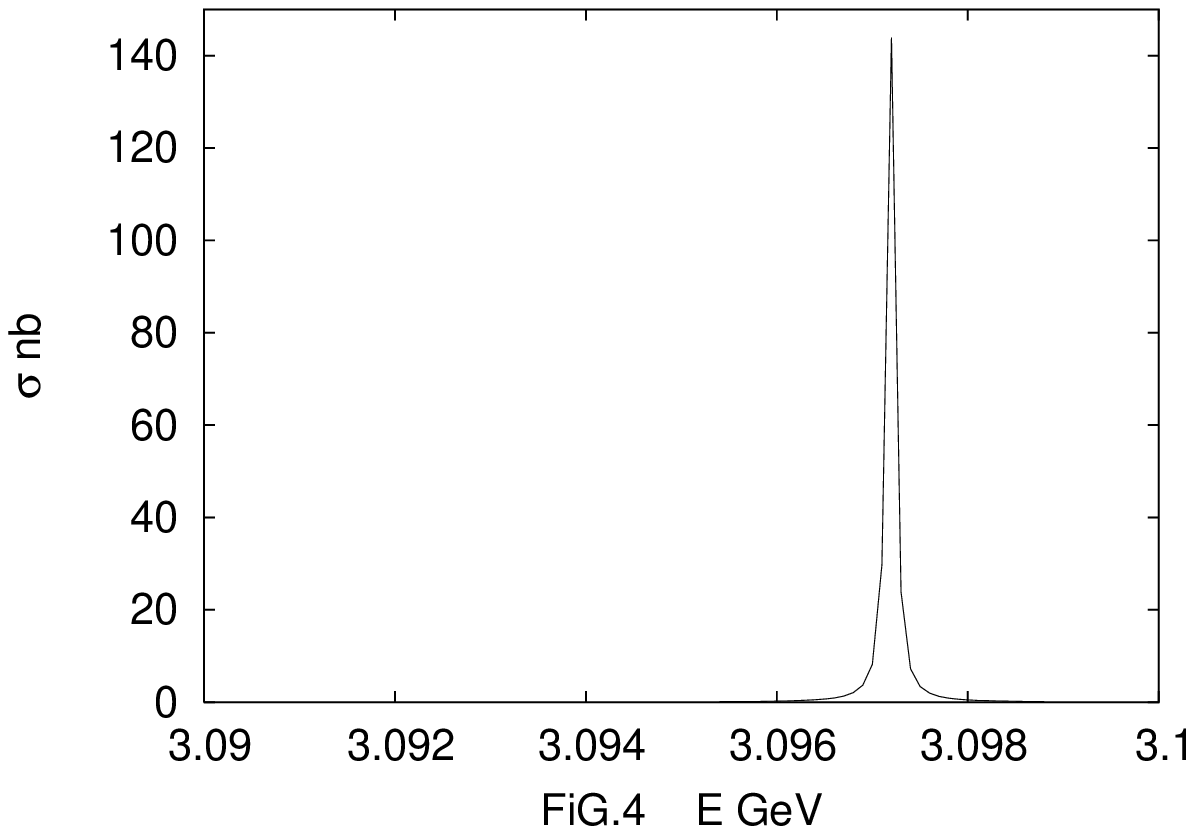}
FIG. 4.
\end{center}
\end{figure}
\begin{figure}
\begin{center}
\psfig{figure=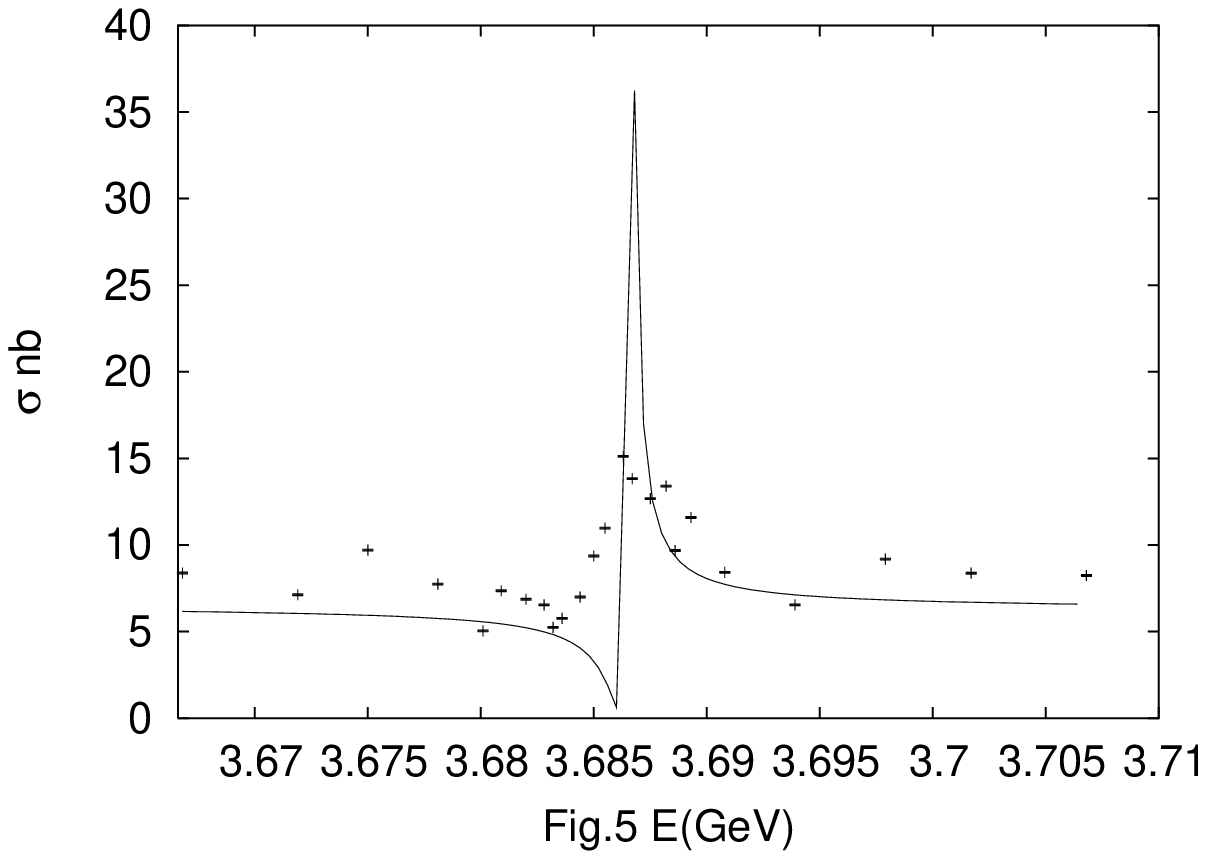}
FIG. 5.
\end{center}
\end{figure}
\begin{figure}
\begin{center}
\psfig{figure=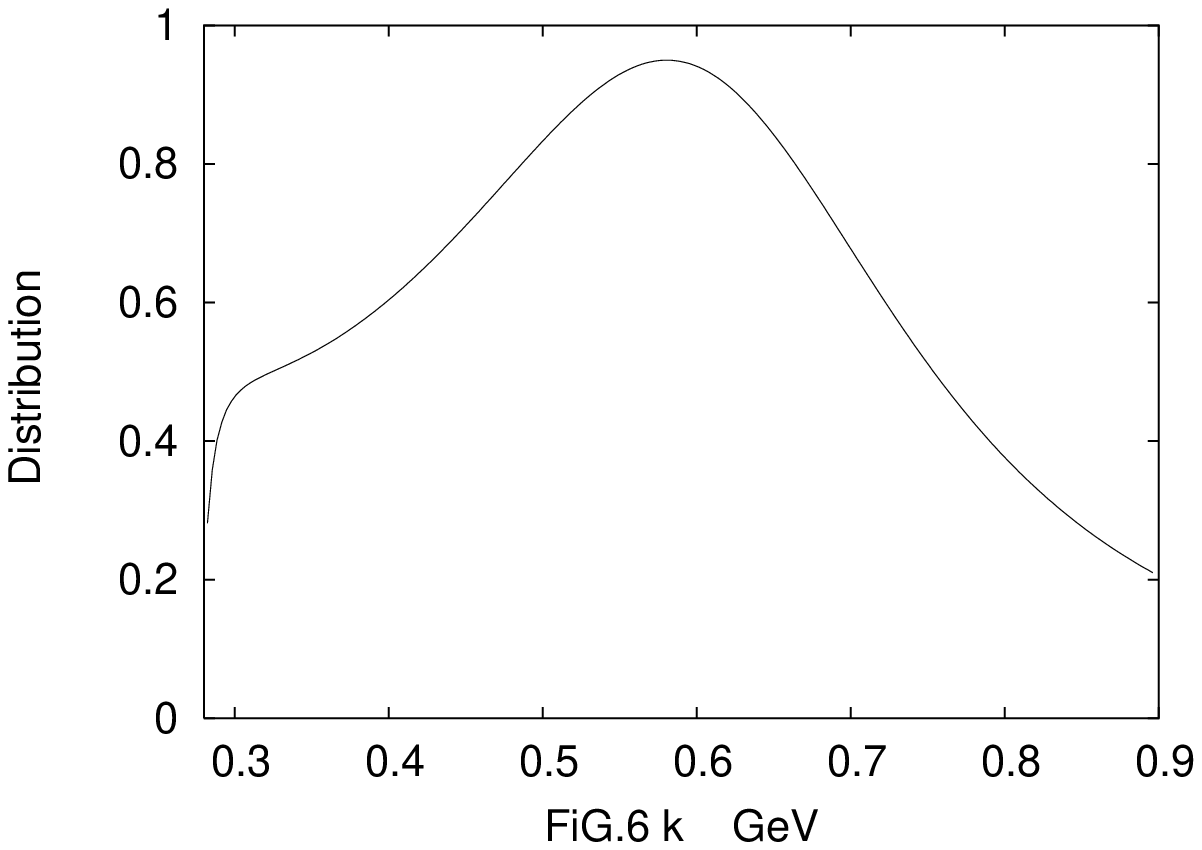}
FIG. 6.
\end{center}
\end{figure}
\begin{figure}
\begin{center}
\psfig{figure=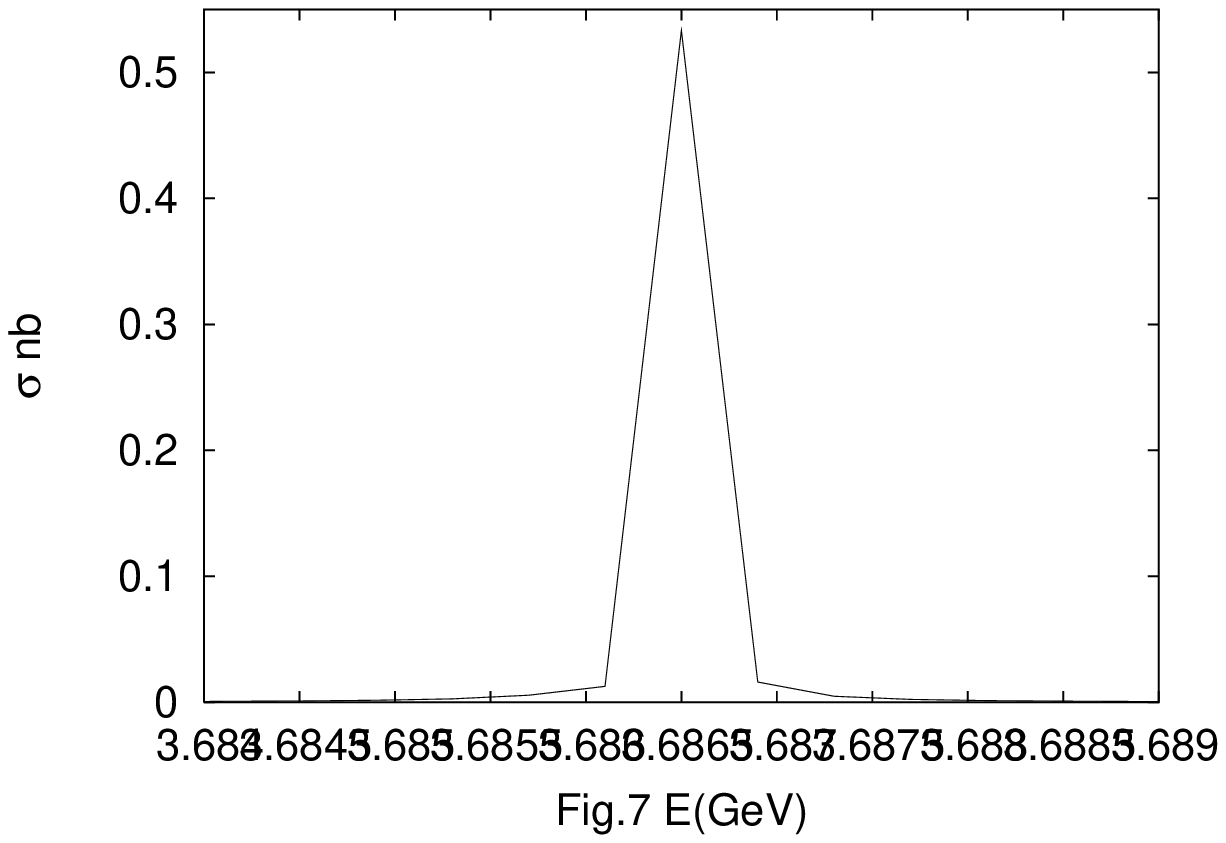}
FIG. 7.
\end{center}
\end{figure}
\begin{figure}
\begin{center}
\psfig{figure=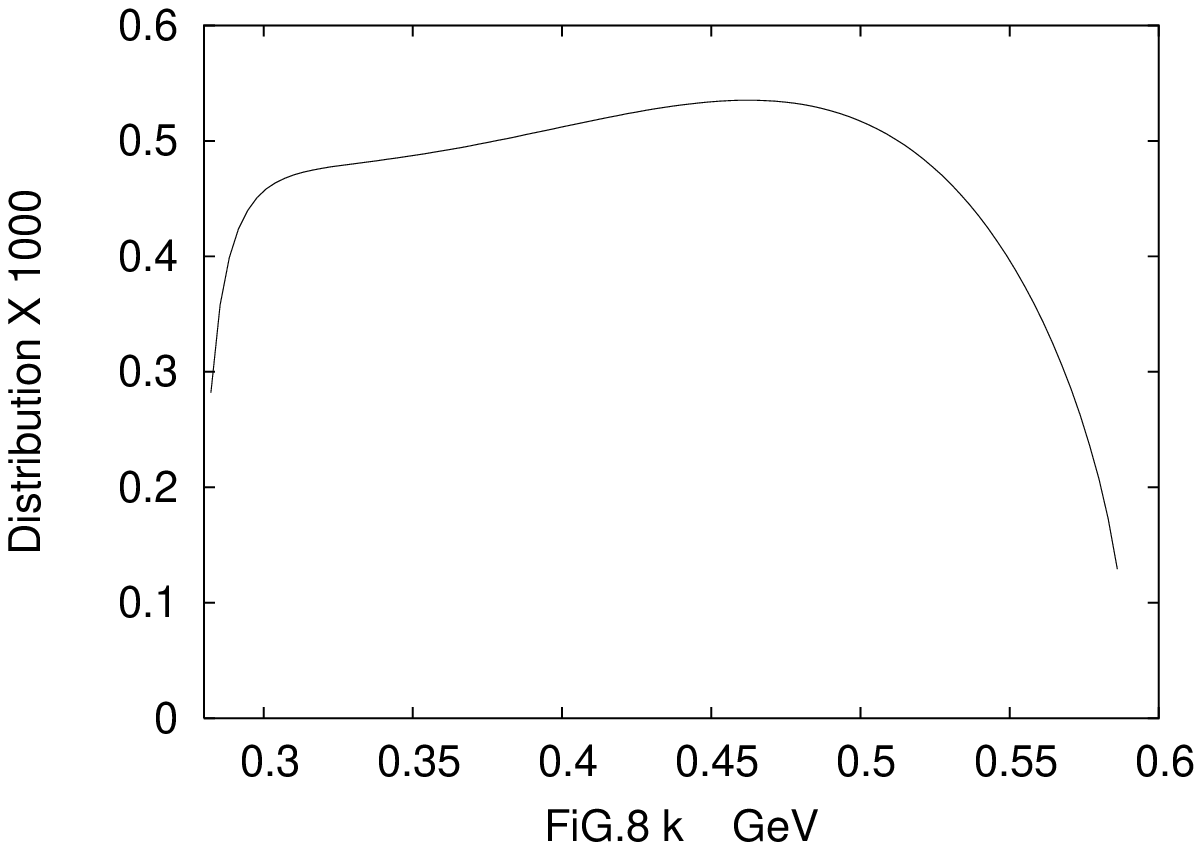}
FIG. 8.
\end{center}
\end{figure}
\begin{figure}
\begin{center}
\psfig{figure=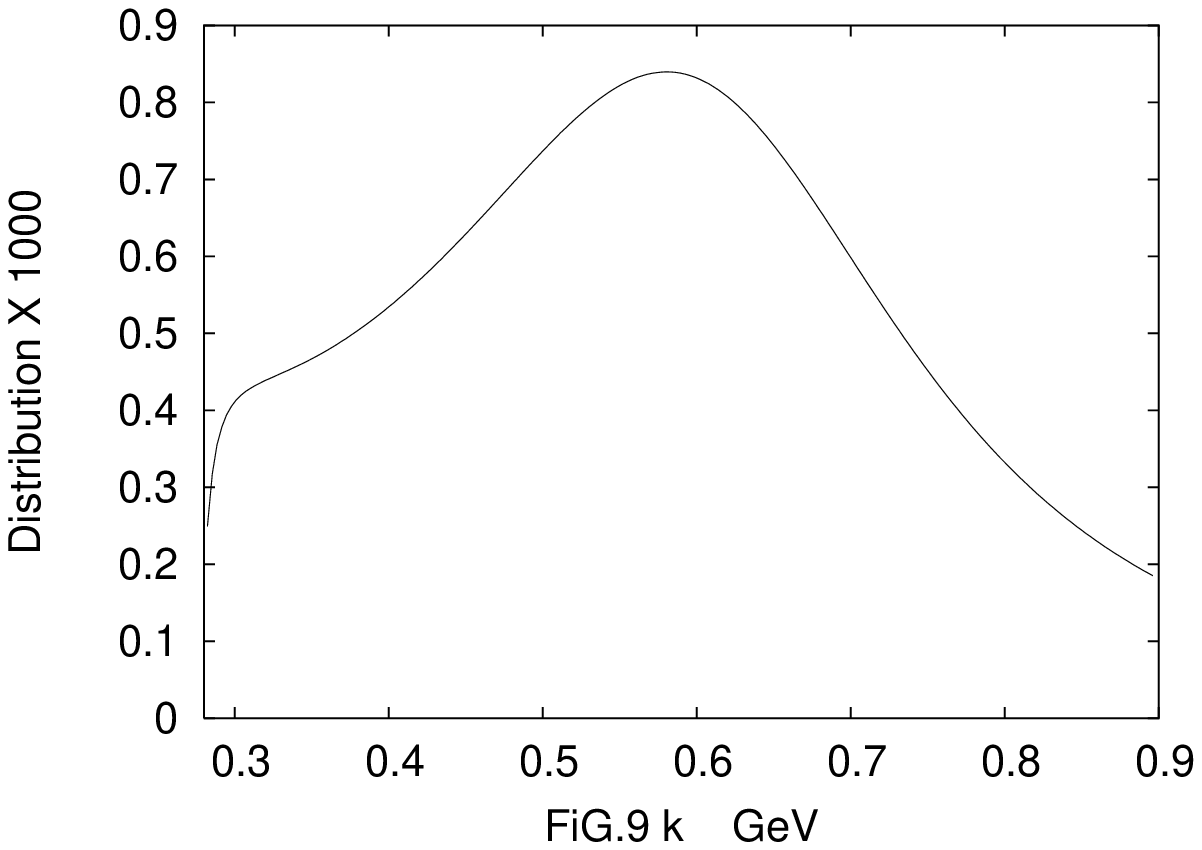}
FIG. 9.
\end{center}
\end{figure}

\end{document}